\documentclass[twocolumn,aps,prl,groupedaddress]{revtex4-2}
\usepackage{bm}
\usepackage{extarrows}
\usepackage{amsfonts}
\usepackage{amsmath}
\usepackage{amssymb}
\usepackage{geometry}
\usepackage{graphicx}
\usepackage{footmisc}
\usepackage{braket}
\usepackage{array}
\usepackage{threeparttable}
\usepackage{bibentry,natbib}
\usepackage{bigstrut}
\usepackage[colorlinks,linkcolor=blue,anchorcolor=blue,citecolor=blue,urlcolor=blue]{hyperref}
\begin{document}

\title{Precision calculation of hyperfine structure of $^{7,9}$Be$^{2+}$ ions}

\author{Xiao-Qiu Qi$^{1,2}$} 
\author{Pei-Pei Zhang$^{2,4,\dag}$}  %~\footnotetext{$\dag$Email address: zhangpei@wipm.ac.cn} 
\author{Zong-Chao Yan$^{3,2}$}
\author{Ting-Yun Shi$^{2}$ }
\author{G. W. F. Drake$^{4}$}
\author{Ai-Xi Chen$^{1}$}
\author{Zhen-Xiang Zhong$^{2}$}

\affiliation {$^1$ Key Laboratory of Optical Field Manipulation of Zhejiang Province and Physics Department of Zhejiang Sci-Tech University, Hangzhou 310018, China}
\affiliation {$^2$ State Key Laboratory of Magnetic Resonance and Atomic and Molecular Physics, Wuhan Institute of Physics and Mathematics, Innovation Academy for Precision Measurement Science and Technology, Chinese Academy of Sciences, Wuhan 430071, China}
\affiliation {$^3$ Department of Physics, University of New Brunswick, Fredericton, New Brunswick, Canada E3B 5A3}
\affiliation {$^4$ Department of Physics, University of Windsor, Windsor, Ontario, Canada N9B 3P4}

\begin{abstract}
The hyperfine structures of the $2\,^3\!S_1$ and $2\,^3\!P_J$ states of the $^7$Be$^{2+}$ and $^9$Be$^{2+}$ ions are investigated within the framework of the nonrelativistic quantum electrodynamics (NRQED). The uncertainties of present hyperfine splitting results of $^9$Be$^{2+}$ are in the order of several tens of ppm, where two orders of magnitude improvement over the previous theory and experiment values has been achieved. The contribution of nuclear electric quadrupole moment to hyperfine splitting of $^7$Be$^{2+}$ has been studied.
A scheme for determining the properties of Be nuclei in terms of Zemach radius or the electric quadrupole moment based on precise spectra is proposed, and it opens a new window for the study of Be nuclei.
\end{abstract}
\date{\today}
\maketitle

\section{Introduction}\label{sec:1}

Light helium and helium-like ions are among simplest atomic systems where theoretical approaches are well advanced to calculate their electronic structures with high precision. However, there are still unsolved interesting problems~\cite{Yan1995,Pachucki2016,Pachucki2017_1,Pachucki2019,Qi2020,Patkos2021a,Patkos2021b}.
Among various theoretical methods,
the nonrelativistic quantum electrodynamics (NRQED) is the most effective approach designed to calculate the electronic structure of light atomic systems~\cite{Pachucki2004,Pachucki2005_1,Pachucki2005_2,Pachucki2017}. For the 
helium $2\,^3\!P_J$ fine-structure, for example, the NRQED-based calculation has achieved a precision of about 1.7~kHz, far exceeding all other theoretical approaches that are based on Dirac-like methods~\cite{Pachucki2010}.
Experimentally, Clausen {\it et al.}~\cite{Clausen2021} have recently reported a much improved new determination of the He $2\,^1\!S$ ionization energy at the level of 32~kHz, which is in good accord with theory. 
However, the derived experimental ionization energies of the $2\,^3\!S$ and $2\,^3\!P$ states are in disagreement with theoretical prediction by $6.5\sigma$ and $10\sigma$, respectively.
${\rm Li}^+$ is very similar to helium with a higher $Z$, and its QED effect is more significant than helium. 
For the $2\,^3\!P_1$-$2\,^3\!P_2$ fine structure interval for example, the contribution from order m$\alpha^6$ and higher in ${\rm Li}^+$ is a factor of 26 larger than for helium~\cite{Guan2020}. The hyperfine structure splittings (hfs) of ${\rm Li}^+$ have been studied in our previous work~\cite{Qi2020} using the NRQED theory. The theoretical uncertainty is reduced to be less than 100~kHz by a complete calculation of all the corrections up to m$\alpha^6$. The so-called Zemach radius, which describes the distribution of magnetic moment inside the nucleus, can be extracted by combining precision measurements~\cite{Guan2020}. The obtained Zemach radius for $^7$Li is in good agreement with previous values, while the value for $^6$Li disagrees with the nuclear physics value~\cite{Yerokhin2008} by more than $6\sigma$, indicating an anomalous nuclear structure for $^6$Li.

For further testing QED effect with low-$Z$ ions, the helium-like $\rm{Be}^{2+}$ is a suitable candidates~\cite{Scholl1993,Yan2008,Wilfried2015}, since the transition wavelength of $2\,^3\!S-2\,^3\!P$ of 372~nm is still close to the visible region. Beryllium has many isotopes $^6$Be-$^{14}$Be~\cite{Labiche1999,Nortershauser2009,Fortune2012,Krieger2012}, including one-neutron halo $^{11}$Be and two-neutron halo $^{14}$Be. 
There are some recent spectral experiments to explore $\rm{Be}$ nuclear structure~\cite{Nortershauser2009,Krieger2012,Drake2013,Nortershauser2015}. 
Puchalski {\it et al.} calculated the hyperfine splittings of $^9$Be using explicitly correlated Gaussian function (ECG), and accurately determined the nuclear electric quadrupole moment~\cite{Puchalski2021a}, although it was inconsistent with the previous value.
The advantage of studying $\rm{Be}^{2+}$ ion rather than neutral Be is that it is a three-body system for which the corresponding QED theory is relatively simpler. 
Compared with helium and ${\rm Li}^+$ ions, the current research on $\rm{Be}^{2+}$ is rare. 
yanyan
In 1993, Scholl {\it et al.} measured the $1s2s^3\!S_1-1s2p^3\!P_J$ transition of $^9\mathrm{Be}^{2+}$ ion by applying fast ion beam laser fluorescence method with an accuracy of $10^{-8}$~\cite{Scholl1993}, which is three orders of magnitude improvement over previous measurements. 
The fine and hyperfine splittings extracted are, respectively, in the order of tens of ppm and $10^{-4}$.
Theoretically, Johnson {\it et al.} in 1997~\cite{Johnson1997} calculated $2\,^3\!P_J$ hfs of $^9\mathrm{Be}^{2+}$ by the relativistic configuration interaction method with only four significant digits. With the development of experimental technology, especially the emergence of new light sources of narrow linewidth in the XUV area~\cite{Jones2005,Arman2012,Zhang2020}, it is now possible to improve the measurement of $\mathrm{Be}^{2+} $ to reach a new accuracy level. 

In this paper, we intend to present a systematic calculation of hfs of the $2\,^3\!S_1$ and $2\,^3\!P_J$ states of the $^{7,9}$Be$^{2+}$ ions by including QED corrections up to $m\alpha^6$ order.
The possibility of determining the Zemach radius and the electric quadrupole moment of a Be isotope based on $\mathrm{Be}^{2+}$ spectroscopy is discussed.
The present paper is organized as follows. 
Sec.~\ref{sec:2} outlines the basic theoretical framework for our calculations. 
Sec.~\ref{sec:3}  details various QED contributions to the hfs of $2\,^3\!S_1$ and $2\,^3\!P_J$ states of the $^{7,9}$Be$^{2+}$.
Finally, discussions and conclusions are given in Sec.~\ref{sec:4}.

\begin{figure}[ht]
\includegraphics[scale=0.5]{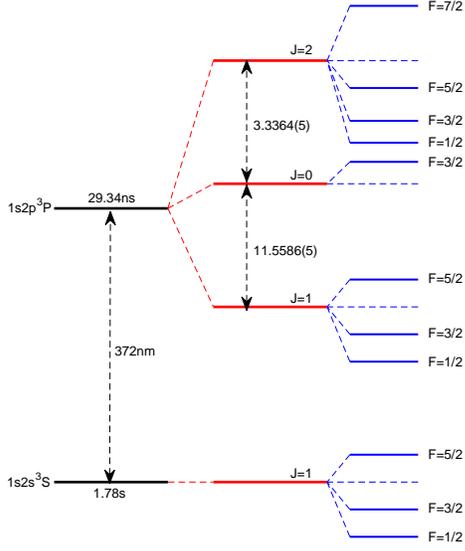}
\caption{Hyperfine energy levels (not drawn to scale) of the $2\,^3\!S_1$ and $2\,^3\!P_J$ states of $^9\mathrm{Be}^{2+}$~\cite{Scholl1993}, in cm$^{-1}$. The nuclear spin of $^9$Be is 3/2.}
\label{fig:1}
\end{figure}

\section{Theoretical method}\label{sec:2}

The NRQED theory for quasidegenerate states is used to calculate fine and hyperfine structure splittings~\cite{Puchalski2009,Pachucki2012,Pachucki2019,Haidar2020}. The theory has been used in calculation of Li$^+$ hyperfine structure~\cite{Qi2020}. Here we simply outline the framework for calculating relativistic and QED corrections to an energy level. Figure~\ref{fig:1} shows the energy level diagram of hfs for $^9\mathrm{Be}^{2+}$ (The diagram for $^7\mathrm{Be}^{2+}$ is similar to $^9\mathrm{Be}^{2+}$ since they have the same nuclear spin 3/2). In order to obtain the energies of the $2\,^3\!S_1$ and $2\,^3\!P_J$ states, we need to diagonalize the effective Hamiltonian $H$ with its matrix elements being
\begin{equation}\label{eq:1}
\begin{aligned}
E_{JJ'}^F\equiv \langle JFM_F|H|J'FM_F \rangle,
\end{aligned}
\end{equation}
where $M_F$ is the projection of the total angular momentum $F$, which can be fixed arbitrarily since the energies are independent of it. For convenience, we treat the $2\,^3P_J$ centroid as the zero level. The above matrix elements Eq.~(\ref{eq:1}) can be expanded in powers of the fine structure constant $\alpha$
\begin{equation}\label{eq:2}
    \begin{aligned}
    E_{JJ'}^F  =&\langle H_{\mathrm{fs}}\rangle_J \delta_{JJ'} + \langle H_{\mathrm{hfs}}^{(4+)}\rangle + \langle H_{\mathrm{hfs}}^{(6)}\rangle \\
    &+ 2\langle H_{\mathrm{hfs}}^{(4)} , [H_{\mathrm{nfs}}^{(4)}+H_{\mathrm{fs}}^{(4)}] \rangle  + \langle H_{\mathrm{hfs}}^{(4)} , H_{\mathrm{hfs}}^{(4)} \rangle \\
    &+ \langle H_{\mathrm{QED}}^{(6)}\rangle + \langle H_{\mathrm{QED}}^{\mathrm{ho}} \rangle + \langle H_{\mathrm{nucl}} \rangle + \langle H_{\mathrm{eqm}} \rangle,
    \end{aligned}
\end{equation}
where $\langle A,B \rangle \equiv \langle A \frac{1}{(E_0-H_0)'} B \rangle$, with $H_0$ and $E_0$ being the nonrelativistic Hamiltonian and its eigenvalue. $H_{\mathrm{fs}}$ is the effective operator that does not depend on the nuclear spin and is responsible for the fine structure splittings~\cite{Pachucki2010,Riis1994}. The other terms in Eq.~(\ref{eq:2}) are the nuclear spin dependent contributions. $H_{\mathrm{hfs}}^{(4+)}$ is the leading-order hyperfine Hamiltonian of $m\alpha^4$, where the superscript `+' means the higher-order terms from the recoil and anomalous magnetic moment effects. $H_{\mathrm{hfs}}^{(6)}$ is the effective operator for the hyperfine  splittings of order $m\alpha^6$. $H_{\mathrm{fs}}^{(4)}$ and $H_{\mathrm{nfs}}^{(4)}$ are the Breit Hamiltonians of order $m\alpha^4$ with and without electron spin. The fifth term in Eq.~(\ref{eq:2}) is the second-order hyperfine correction, which contributes to the isotope shift, fine and hyperfine splittings. $H_{\mathrm{QED}}^{(6)}$ and $H_{\mathrm{QED}}^{\mathrm{ho}}$ are the two effective operators for the QED corrections of order $m\alpha^6$ and higher $\sim m\alpha^7$. Finally, $H_{\mathrm{nucl}}$ and $H_{\mathrm{eqm}}$ represent the nuclear effects due to the Zemach radius and the nuclear electric quadrupole moment.

We solve the eigenvalue problem of $H_0$ variationally in Hylleraas coordinates.
The relativistic and QED corrections as well as the corrections due to nuclear structure are evaluated perturbatively.
The Hylleraas basis set~\cite{Zhang2015} is constructed according to
\begin{equation}\label{eq:3}
\begin{aligned}
\psi_{\ell mn}(\vec{r}_1,\vec{r}_2) = r^{\ell}_1r^m_2r^n e^{-\alpha r_1-\beta r_2-\gamma r} \mathcal{Y}^{LM}_{{\ell}_1{\ell}_2}(\hat{r}_1,\hat{r}_2),
\end{aligned}
\end{equation}
where $\vec{r}=\vec{r}_1-\vec{r}_2$ and $\mathcal{Y}^{LM}_{{\ell}_1{\ell}_2}(\hat{r}_1,\hat{r}_2)$ is the vector coupled product of spherical harmonics for the electrons. In order to deal with the nonrelativistic finite nuclear mass effect, according to whether the mass polarization operator is explicitly included in the nonrelativistic Hamiltonian, two different types of wave functions can be generated. 
For $\langle H_{\mathrm{hfs}}^{(4+)}\rangle$, $\langle H_{\mathrm{QED}}^{(6)}\rangle$, $\langle H_{\mathrm{QED}}^{\mathrm{ho}} \rangle$, $\langle H_{\mathrm{nucl}} \rangle$, and $\langle H_{\mathrm{eqm}} \rangle$, we use the wave functions with the mass polarization, whereas for other terms we use the wave functions corresponding to the infinite nuclear mass limit. The coupling of intermediate states of different symmetries should be included in the second-order terms, where some singular integrals need to be handled by including more singular terms in the intermediate states~\cite{Drake2002}. The necessary
angular momentum operators, which can be evaluated analytically~\cite{Qi2020},
are~\cite{Pachucki2012} $S^iL^i$, $I^iL^i$, $I^iS^i$, $\{S^iS^j\}\{L^iL^j\}$, $I^iS^j\{L^iL^j\}$, $I^iL^j\{S^iS^j\}$, $\{I^iI^j\}\{S^iS^j\}$, $\{I^iI^j\}\{L^iL^j\}$, $\{I^iI^j\}L^iS^j$, and $\{I^iI^j\}[\{S^mS^n\}\{L^kL^l\}]^{ij}$, where $\{S^iS^j\}\equiv \frac{1}{2}S^iS^j + \frac{1}{2}S^jS^i - \frac{1}{3}\vec{S}^2\delta^{ij}$ and the summation over the repeated indices is assumed.

\section{The hfs of $2\,^3\!S_1$ and $2\,^3\!P_J$ states}\label{sec:3}

The hfs operators responsible for relativistic and QED corrections to the $2\,^3\!S_1$ and $2\,^3\!P_J$ states of helium-like ions are defined in our previous paper~\cite{Qi2020}.
The first-order perturbation results of $m\alpha^4$ and $m\alpha^6$ corrections are listed in Table~\ref{tab:1}.

\begin{table}[ht]
    \caption{\label{tab:1} Expectation values for the $2\,^3\!S_1$ and $2\,^3\!P_J$ states of $^7$Be$^{2+}$ and $^9$Be$^{2+}$. The listed numerical values are uncertain only at the last digits. In atomic units.}
    \begin{ruledtabular}
    \begin{tabular}{lcccc}
     \multicolumn{1}{c}{State}
    &\multicolumn{1}{c}{Operator}
    &\multicolumn{1}{c}{$^7\mathrm{Be}^{2+}$}
    &\multicolumn{1}{c}{$^9\mathrm{Be}^{2+}$} \\
    \hline
    $2\,^3\!S_1$ &$4\pi \delta^3(\vec{r}_1)$                 &137.731960       &137.739110   \\
    $          $ &$K'$                                       &--157.232037     &--157.232037 \\
    $2\,^3\!P_J$ &$4\pi \delta^3(\vec{r}_1)$                 &126.232881       &126.239606   \\
    $          $ &$(\vec{r}_1 \times \vec{p}_1)/r_1^3$       &1.814143         &1.814146     \\
    $          $ &$(\vec{r}_1 \times \vec{p}_2)/r_1^3$       &--2.602150       &--2.602181   \\
    $          $ &$(\delta^{ij} - 3r_1^ir_1^j/r_1^2)/r_1^3$  &--0.671762       &--0.671769   \\
    $          $ &$K'$                                       &55.3670854       &55.3670854   \\
    $          $ &$\vec{K}$                                  &--145.86034      &--145.86034  \\
    $          $ &$\hat{K}$                                  &--82.07829       &--82.07829   \\
    \end{tabular}
    \end{ruledtabular}
\end{table}

The second-order corrections of $m\alpha^6$ can be divided into several parts according to the symmetries of the intermediate states. 
For the $2\,^3\!S_1$ state, the intermediate states are $^3\!S$, $^3\!P$, and $^3\!D$. 
And for $2\,^3\!P_J$ state the intermediate states are $^3P$, $^1P$, $^3D$, $^1D$, and $^1F$. 
Numerical results of various operators for the radial parts are presented in Table~\ref{tab:2}. 
Since the second-order hyperfine correction $\langle H_{\mathrm{hfs}}^{(4)} , H_{\mathrm{hfs}}^{(4)} \rangle$ is divergent, we calculate only the dominant contribution from the $2\,^1\!P_1$ intermediate state. 
It should be noted that the uncertainty of $\langle P_A,P_A \rangle^\circ$ in Table~\ref{tab:2} is only computational. 
We also use the method in Ref.~\cite{Pachucki2012} to estimate the uncertainty due to this approximation, {\it i.e.}, calculating the second-order perturbation for the operator $\langle P'_A,P'_A \rangle$, and taking the difference between $\langle P_A,P_A \rangle^\circ$ and $\langle P'_A,P'_A \rangle$ as the uncertainty, which is 10000~a.u. for $2\,^3\!S_1$ and 15000~a.u. for $2\,^3\!P_J$ respectively.

\begin{table}[ht]
    \caption{\label{tab:2} Second-order matrix elements for all possible intermediate states connected to the $2\,^3\!S_1$ and $2\,^3\!P_J$ states. The listed numerical values are uncertain only at the last digits when not given explicitly. In atomic units.}
     \begin{ruledtabular}
     \begin{tabular}{lccccc}
     \multicolumn{1}{c}{State}
    &\multicolumn{1}{c}{Symmetry}
    &\multicolumn{1}{c}{$\langle A, B \rangle$}
    &\multicolumn{1}{c}{Value} \\
    \hline
    $2\,^3\!S_1$   &$^3\!S$&$\langle P',G' \rangle$                       & 4592.8              \\
                   &$^3\!P$&$\langle \vec{P},\vec{G} \rangle$             & 0.140               \\
                   &$^3\!D$&$\langle \hat{P},\hat{G} \rangle$             & 0.87                \\
                   &$^1\!S$&$\langle P_A,P_A \rangle^\circ$               & --839249.282        \\
                   &       &$\langle P'_A,P'_A \rangle    $               & --848800(200)       \\
    $2\,^3\!P_J$   &$^3\!P$&$\langle P',G' \rangle$                       & 4024.6(5)           \\
                   &       &$\langle \vec{P},G \rangle$                   & 86.9(5)             \\
                   &       &$\langle \hat{P},G \rangle$                   & 40(5)               \\
                   &       &$\langle P,\vec{G} \rangle$                   & 26.678              \\
                   &       &$\langle \vec{P},\vec{G} \rangle$             & --64.1              \\
                   &       &$\langle \hat{P},\vec{G} \rangle$             & --39.8427           \\
                   &       &$\langle P,\hat{G} \rangle$                   & 6.714(5)            \\
                   &       &$\langle \vec{P},\hat{G} \rangle$             & 19.323              \\
                   &       &$\langle \hat{P},\hat{G} \rangle$             & 8.882               \\
                   &$^1\!P$&$\langle P_A,\vec{G}_A \rangle$               & 9054.88(5)          \\
                   &       &$\langle P_A,\vec{G}_A \rangle^\circ$         & 9021.158            \\
                   &       &$\langle \hat{P}_A,\vec{G}_A \rangle$         & --176.7(5)          \\
                   &       &$\langle \hat{P}_A,\vec{G}_A \rangle^\circ$   & --139.044           \\
                   &$^3\!D$&$\langle \vec{P},\vec{G} \rangle$             & 0.044               \\
                   &       &$\langle \hat{P},\vec{G} \rangle$             & --0.0149824         \\
                   &       &$\langle \vec{P},\hat{G} \rangle$             & 1.563(5)            \\
                   &       &$\langle \hat{P},\hat{G} \rangle$             & --0.0688            \\
                   &$^1\!D$&$\langle \hat{P}_A,\vec{G}_A \rangle$         & 0.348(5)            \\
                   &$^3\!F$&$\langle \hat{P},\hat{G} \rangle$             & 0.628(5)            \\
                   &$^1\!P$&$\langle P_A,P_A \rangle^\circ$               & --1785103.485       \\
                   &       &$\langle P'_A,P'_A \rangle$                   & --1797300(200)      \\
                   &       &$\langle P_A,\hat{P}_A \rangle^\circ$         & 27527.773           \\
                   &       &$\langle \hat{P}_A,\hat{P}_A \rangle^\circ$   & --2547.006          \\
    \end{tabular}
    \end{ruledtabular}
\end{table}

We calculate the hfs of the $2\,^3\!S_1$ and $2\,^3\!P_J$ states using the values in Tables~\ref{tab:1} and \ref{tab:2}. 
Since the contribution from the $1s$ electron dominates higher-order QED correction, 
the assumption that $H_{\mathrm{QED}}^{\mathrm{ho}} (1s2p) \simeq H_{\mathrm{QED}}^{\mathrm{ho}} (1s)$ 
is adopted for the hfs calculation of the $2\,^3\!P_J$ state, while the $H_{\mathrm{QED}}^{\mathrm{ho}} (1s2s)$ of $2\,^3\!S_1$ state 
is approximated by the weighted average of $H_{\mathrm{QED}}^{\mathrm{ho}} (1s)$ and $H_{\mathrm{QED}}^{\mathrm{ho}} (2s)$. 
The uncertainty of the correction $H_{\mathrm{QED}}^{\mathrm{ho}}$ is estimated as 20\% of its contribution. 
According to Eq.~(\ref{eq:2}), the hfs calculation of the $2\,^3\!P_J$ state requires the results of the fine structure splittings, 
which are $\langle H_{fs}\rangle_{J=0}=(8f_{01}+5f_{12})/9$, $\langle H_{fs}\rangle_{J=1}=(-f_{01}+5f_{12})/9$, 
and $\langle H_{fs}\rangle_{J=2}=(-f_{01}-4f_{12})/9$, relative to the $2\,^3\!P_J$ centroid, 
with $f_{01} =11.5586(5)$~cm$^{-1}$ and $f_{12} =-14.8950(4)$~cm$^{-1}$ for $^9$Be$^{2+}$~\cite{Scholl1993}. 
The fine structure splittings of $^7$Be$^{2+}$ are obtained by changing the reduced mass accordingly
{\it i.e.}, $f_{01} =11.558(2)$~cm$^{-1}$ and $f_{12} =-14.895(2)$~cm$^{-1}$.
For $^7$Be$^{2+}$ and $^9$Be$^{2+}$,
the magnetic moments are $-1.39928(2)~\mu_N$~\cite{Okada2008} and
$-1.177432(3)~\mu_N$~\cite{Nortershauser2009}, and the nuclear electric quadrupole moments $-6.11~\rm{fm}^2$ (the theoretical result from ~\cite{Koji2001}) and $5.350(14)~\rm{fm}^2$~\cite{Puchalski2021a}, respectively. 
The contributions of Zemach radii are at $-615(8)$~ppm~\cite{Puchalski2021a} for $^9$Be$^{2+}$ and $-521(16)$~ppm for $^7$Be$^{2+}$. 
This contribution of $^7$Be$^{2+}$ is calculated by $-2ZR_{\rm{em}}/a_0$, where $R_{\rm{em}}=4R_{\rm{e}}/\sqrt{3\pi}$ (Gaussian distributions) and the nuclear charge radius $R_{\rm{e}}$ is 2.647(17) fm~\cite{Nortershauser2009}.
The hfs of $2\,^3\!S_1$ and $2\,^3\!P_J$ states can be obtained by diagonalizing the matrix in Eq.~(\ref{eq:2}) 
and the results relative to the $2\,^3\!S_1$ and $2\,^3\!P_J$ centroids are listed in Table~\ref{tab:3}.

\begin{table}[ht]
    \scriptsize\caption{\label{tab:3} Theoretical results for individual $2\,^3\!S^F_1$ and $2\,^3\!P^F_J$ levels in $^7$Be$^{2+}$ and $^9$Be$^{2+}$, relative to the $2\,^3\!S_1$ and $2\,^3\!P_J$ centroid respectively, where the first error in $2^3P$ state is due to the fine structure and the second error is due to the hyperfine structure, in cm$^{-1}$.}
    \begin{ruledtabular}
    \begin{tabular}{lcccc}
     \multicolumn{1}{l}{State}
    &\multicolumn{1}{l}{$(J,F)$}
    &\multicolumn{1}{c}{$^7\mathrm{Be}^{2+}$}
    &\multicolumn{1}{c}{$^9\mathrm{Be}^{2+}$}  \\
    \hline
    $2^3S$ &$(1,1/2)$&\,\,\,0.68251(1)          &\,\,\,0.574282(6)          \\
           &$(1,3/2)$&\,\,\,0.27300(1)          &\,\,\,0.229708(3)          \\
           &$(1,5/2)$&--0.40950(1)              &--0.344566(4)              \\
    $2^3P$ &$(2,1/2)$&\,\,\,5.90767(100)(1)     &\,\,\,5.817172(190)(4)     \\
           &$(2,3/2)$&\,\,\,5.72041(100)(1)     &\,\,\,5.658805(190)(3)     \\
           &$(2,5/2)$&\,\,\,5.40467(100)(1)     &\,\,\,5.392683(190)(1)     \\
           &$(2,7/2)$&\,\,\,4.95513(100)(1)     &\,\,\,5.015548(190)(3)     \\
           &$(0,3/2)$&\,\,\,2.01174(200)(1)     &\,\,\,2.008479(500)(1)     \\
           &$(1,1/2)$&--9.23556(110)(1)         &--9.287087(230)(3)         \\
           &$(1,3/2)$&--9.44648(110)(1)         &--9.461891(230)(1)         \\
           &$(1,5/2)$&--9.75933(110)(1)         &--9.727037(230)(2)         \\
    \end{tabular}
    \end{ruledtabular}
\end{table}

For the $2\,^3\!P_J$ states, the $2\,^1\!P_1-2\,^3\!P_1$ mixing effect should be taken into consideration carefully.
Here we follow two methods used in our previous calculation~\cite{Qi2020}. Method 1. Do an exact diagonalization only within the $2\,^3\!P_J$ manifold and treat the $2\,^1\!P_1-2\,^3\!P_1$ mixing effect by perturbation theory up to second order. Method 2. Extend the $2\,^3\!P_J$ manifold by including the $2\,^1\!P_1$ state and do an exact diagonalization of the extended matrix. Both the methods only include the relativistic correction of order m$\alpha^4$.
The second-order matrix elements involving the intermediate state $2\,^1\!P_1$ and the hyperfine structure coefficients~\cite{Riis1994} for the $2\,^1\!P_1$ and $2\,^3\!P_J$ states are listed in Tables~\ref{tab:2} and \ref{tab:4}, as inputs for applying Methods 1 and 2. The hfs of $2\,^3\!P_J$ are evaluated using these two methods 
and the results are presented in Table~\ref{tab:5}. The modification of the mixing effect alters the hyperfine intervals $(1,1/2)-(1,3/2)$ and $(1,3/2)-(1,5/2)$  
by 0.000322~cm$^{-1}$ and 0.000516~cm$^{-1}$ for $^9$Be$^{2+}$, whereas for $^7$Be$^{2+}$ they are 0.00038~cm$^{-1}$ and 0.00061~cm$^{-1}$, respectively.
These shifts are about three orders of magnitude larger than that of $^7$Li$^+$.
Our final results of $2\,^3\!P_J$ hfs for $^7$Be$^{2+}$ and $^9$Be$^{2+}$ are shown in Tables~\ref{tab:6} and Table~\ref{tab:7}.

\begin{table}[ht]
\caption{\label{tab:4} Calculated values of $^7$Be$^{2+}$ and $^9$Be$^{2+}$ hfs coefficients for the $2\,^1\!P_1$ and $2\,^3\!P_J$ states, in cm$^{-1}$. These coefficients are defined in Eqs.~(10)-(12) of Ref.~\cite{Riis1994}. The listed numerical values are uncertain only at the last digits.}
    \begin{ruledtabular}
    \begin{tabular}{lcccc}
     \multicolumn{1}{c}{Coefficient}
    &\multicolumn{1}{c}{$^7\mathrm{Be}^{2+}$}
    &\multicolumn{1}{c}{$^9\mathrm{Be}^{2+}$} \\
    \hline
     $C^{(0)}_{1,1}$     &--0.24990          &--0.210283       \\
     $C^{(0)}_{1,0}$     &--0.25126          &--0.211428       \\
     $D^{(0)}_{1}$       &--0.00440          &--0.003701       \\
     $D^{(0)}_{0}$       &--0.00317          &--0.002663       \\
     $E^{(0)}_{1,1}$     &\,\,\,0.00112      &\,\,\,0.000938   \\
     $E^{(0)}_{1,0}$     &\,\,\,0.00097      &\,\,\,0.000815   \\
    \end{tabular}
    \end{ruledtabular}
\end{table}

\begin{table}[ht]
    \caption{\label{tab:5} Hyperfine splittings in $2\,^3\!P_J$ of $^7$Be$^{2+}$ and $^9$Be$^{2+}$, in cm$^{-1}$. Only the relativistic correction of order m$\alpha^4$ is included. The listed numerical values are uncertain only at the last digits.}
    \begin{ruledtabular}
    \begin{tabular}{cccccc}
                  &$(J,F)-(J',F')$    &Method 1  &Method 2   &Difference       \\
    \hline
    $^7$Be$^{2+}$ &$(2,1/2)-(2,3/2)$  &0.18729    &0.18729     &               \\
                  &$(2,3/2)-(2,5/2)$  &0.31552    &0.31552     &               \\
                  &$(2,5/2)-(2,7/2)$  &0.44873    &0.44872     &--0.00001      \\
                  &$(1,1/2)-(1,3/2)$  &0.21031    &0.21069     &\,\,\,0.00038  \\
                  &$(1,3/2)-(1,5/2)$  &0.31267    &0.31328     &\,\,\,0.00061  \\
    $^9$Be$^{2+}$ &$(2,1/2)-(2,3/2)$  &0.157960  &0.157964     &\,\,\,0.000004 \\
                  &$(2,3/2)-(2,5/2)$  &0.265658  &0.265660     &\,\,\,0.000002 \\
                  &$(2,5/2)-(2,7/2)$  &0.376899  &0.376891     &--0.000008     \\
                  &$(1,1/2)-(1,3/2)$  &0.174864  &0.175186     &\,\,\,0.000322 \\
                  &$(1,3/2)-(1,5/2)$  &0.264669  &0.265185     &\,\,\,0.000516 \\                 
    \end{tabular}
    \end{ruledtabular}
\end{table}

\begin{table*}[ht]
    \scriptsize\caption{\label{tab:6} Theoretical hyperfine intervals in the $2\,^3\!S_1$ and $2\,^3\!P_J$ states of $^7$Be$^{2+}$ with quadrupole moment $Q_\mathrm{d}=-6.11~\rm{fm}^2$. The listed numerical values are uncertain only at the last digits.}
    \begin{ruledtabular}
    \begin{tabular}{ccccccccc}
    $\rm{State}$  &$(J,F)-(J',F')$     &$E_{\rm{a}}$ &$10^5 X$             &$10^6 \delta X$     &$\eta$        &$E_\mathrm{d}$  \\
                  &                    &cm$^{-1}$    &cm$^{-1}/\rm{fm}^2$  &cm$^{-1}/\rm{fm}^2$ &ppm           &cm$^{-1}$       \\
    \hline                          
    $2\,^3\!P_2$  &$(2,1/2)-(2,3/2)$   &0.18751      &\,\,\,4.12051        &\,\,\,0.30           &1354          &0.18726         \\
    $ $           &$(2,3/2)-(2,5/2)$   &0.31591      &\,\,\,2.94322        &--2.04               &\,\,\,530     &0.31574         \\
    $ $           &$(2,5/2)-(2,7/2)$   &0.44928      &--4.12051            &\,\,\,1.02           &\,\,\,546     &0.44953         \\
    $2\,^3\!P_1$  &$(1,1/2)-(1,3/2)$   &0.21097      &--5.29780            &--0.43               &1544          &0.21130         \\
    $ $           &$(1,3/2)-(1,5/2)$   &0.31365      &\,\,\,2.94322        &\,\,\,2.17           &\,\,\,616     &0.31346         \\
    $2\,^3\!S_1$  &$(1,1/2)-(1,3/2)$   &0.40951      &                     &                     &              &0.40951         \\
                  &$(1,3/2)-(1,5/2)$   &0.68250      &                     &                     &              &0.68250         \\
    \end{tabular}
    \end{ruledtabular}
\end{table*}

\begin{table}[ht]
    \scriptsize\caption{\label{tab:7} Experimental and theoretical hyperfine intervals in the $2\,^3\!S_1$ and $2\,^3\!P_J$ states of $^9$Be$^{2+}$, in cm$^{-1}$.}
    \begin{ruledtabular}
    \begin{tabular}{ccccccccc}
     \multicolumn{1}{c}{}
    &\multicolumn{1}{c}{}
    &\multicolumn{1}{c}{Experiment}
    &\multicolumn{2}{c}{Theory}\\
    \cline{3-3} \cline{4-5}
     $\rm{State}$  &$(J,F)-(J',F')$     &Scholl {\it et al.}   &Johnson {\it et al.}   & This work          \\
                   &                    &\cite{Scholl1993}     &\cite{Johnson1997}     &                    \\
    \hline
     $2\,^3\!P_2$  &$(2,1/2)-(2,3/2)$   &0.1585(10)            &0.1581                 &0.158371(7)         \\
     $ $           &$(2,3/2)-(2,5/2)$   &0.2659(11)            &0.2659                 &0.266123(4)         \\
     $ $           &$(2,5/2)-(2,7/2)$   &0.3768(14)            &0.3773                 &0.377128(4)         \\
     $2\,^3\!P_1$  &$(1,1/2)-(1,3/2)$   &0.1751(10)            &0.1754                 &0.175126(4)         \\
     $ $           &$(1,3/2)-(1,5/2)$   &0.2654(10)            &0.2654                 &0.265662(3)         \\
     $2\,^3\!S_1$  &$(1,1/2)-(1,3/2)$   &0.3448(10)            &                       &0.344574(9)         \\
                   &$(1,3/2)-(1,5/2)$   &0.5740(11)            &                       &0.574275(6)         \\
    \end{tabular}
    \end{ruledtabular}
\end{table}

\section{Discussion and conclusion}\label{sec:4}
The radioactive $^7$Be is a special atomic nucleus whose magnetic moment cannot be obtained by the $\beta\gamma$-NMR method, and optical spectroscopy is the only method to measure the nuclear moment. Although Okada {\it et al.}~\cite{Okada2008} determined the magnetic dipole moment of $^7$Be to high accuracy, its charge radius has not been determined until now. In addition, there is no published value for the nuclear electric quadrupole moment of $^7$Be. Fortunately, although $^7\mathrm{Be}$ is not stable, its half-life is about 53 days, which is helpful for the experimental measurement of $^7$Be. Since the quadrupole moment $Q_\mathrm{d}$ of $^7$Be has not been measured and the results obtained by theoretical calculations differ noticeably from each other (--6.11~$\rm{fm}^2$~\cite{Koji2001}, --5.50(48)~$\rm{fm}^2$ and --4.68(28)~$\rm{fm}^2$ ~\cite{Forssen2009}), these values are not yet conclusive. Here we study the contribution of the quadrupole moment $Q_\mathrm{d}$ to hfs of $^7$Be$^{2+}$ by ignoring its higher-order nonlinear correction,
\begin{equation}\label{eq:4}
\begin{aligned}
E_{\rm{d}} = E_{\rm{a}}+Q_{\rm{d}} (X+\delta X),
\end{aligned}
\end{equation}
where $E_{\rm{d}}$ and $E_{\rm{a}}$ represent the hfs obtained by diagonalization of Eq.~(\ref{eq:2}) with and without the contribution of $Q_{\rm{d}}$ ($\langle H_{\mathrm{eqm}} \rangle$ term). $X$ and $\delta X$ are tow linear coefficients independent of $Q_{\rm{d}}$, where $X$ is obtained from the diagonal element of the $\langle H_{\mathrm{eqm}} \rangle$ term, and $\delta X$ comes from the linear correction caused by the $\langle H_{\mathrm{eqm}} \rangle$ term included in the diagonalization process. In order to reflect the sensitivity of the transitions to $Q_{\rm{d}}$ intuitively, one can define the relative accuracy,
\begin{equation}\label{eq:5}
\begin{aligned}
\eta =\left\lvert \frac{Q_{\rm{d}}(X+\delta X)}{E_{\rm{d}}} \right\rvert.
\end{aligned}
\end{equation}
In other words, the $\eta$ is the precision required to detect the contribution of $Q_{\rm{d}}$ in the experiment. Using the theoretical value $Q_\mathrm{d}=-6.11~\rm{fm}^2$ chosen in the Ref.~\cite{Puchalski2009}, we calculated the results as shown in Table~\ref{tab:6}. The results show that the transitions $(1,1/2)-(1,3/2)$ and $(2,1/2)-(2,3/2)$ are more suitable to be used to determine the $Q_{\rm{d}}$. According to the theoretical values, the value of $Q_d$ is likely to exist between $-7 \; \rm{fm}^2$ and $-4 \; \rm{fm}^2$. In this range, once the experiment reaches the same accuracy as the theory, the result of the $Q_{\rm{d}}$ can be determined according to the Eq.~(\ref{eq:5}), and the accuracy can have two significant digits.

Table~\ref{tab:7} lists the experimental and theoretical hyperfine intervals in the $2\,^3\!P_J$ state of $^9$Be$^{2+}$, where the uncertainties are mainly due to the $m\alpha^7$ contribution and the nuclear structure.
It is worth noting that these theoretical uncertainties are propagated only from the errors displayed in Table~\ref{tab:3}. 
The uncertainty from the fine structure is canceled for the same-$J$ transitions. 
Table~\ref{tab:7} also shows the measured results obtained through weighted average of all the values in Ref.~\cite{Scholl1993}, 
and the only available theoretical values of Johnson \emph{et al.}~\cite{Johnson1997} for the $2\,^3\!P_J$ state.
Our results are in good agreement with these previous values and are about two orders of magnitude more precise.
Our theoretical calculation has reached the level of ten or so ppm, which is sensitive to some of the major nuclear electromagnetic structure effect. 

In summary, we have studied the hfs of the $2\,^3\!S_1$ and $2\,^3\!P_J$ states of the $^7$Be$^{2+}$ and $^9$Be$^{2+}$ ions, 
including the relativistic and QED corrections up to order $m\alpha^6$. 
The $2\,^1\!P_1-2\,^3\!P_1$ single-triple mixing effect has been treated rigorously. Compared to Li$^{+}$, the $2\,^1\!P_1-2\,^3\!P_1$ mixing effect is about three orders of magnitude larger, indicating that this procedure becomes more and more essential with increasing $Z$. 
The uncertainties of present calculations are in the order of tens of ppm for $^9$Be$^{2+}$, 
mainly from the error of $m\alpha^7$ and nuclear contribution (the Zemach radius). The results for the hfs of the $2\,^3\!S_1$ and $2\,^3\!P_J$ states have been improved by two orders of magnitude. 
The contribution of nuclear electric quadrupole moment to the hyperfine splittings of $^7$Be$^{2+}$ has also been studied.
In order to observe the influence of $Q_\mathrm{d}$, the precision of experimental measurements on hfs needs to be better than $10^{-4} \; \rm{cm}^{-1}$. If the experiments reach the same accuracy as the present theoretical value, the two significant digits of $Q_\mathrm{d}$ can be determined. Our results may stimulate further experimental activities to explore Be nuclear structure.

$\dag$Email address: zhangpei@wipm.ac.cn

\bibliography{refx}

\end{document}